# SECURE SUPERVISED LEARNING-BASED SMART HOME AUTHENTICATION FRAMEWORK

K. Swapna Sudha[1], N. Jeyanthi[2], and Celestine Iwendi[3]

[1]Research Scholar, School of Computer Science Engineering and Information Systems, Vellore Institute of Technology, Vellore-14, India
[2]Professor, School of Computer Science Engineering and Information Systems, Vellore Institute of Technology, Vellore-14, India
[3]Research Coordinator, School of Creative Technologies at the University of Bolton.

## ABSTRACT

*The Smart home possesses the capability of facilitating home services to their users with the systematic advance in The Internet of Things (IoT) and information and communication technologies (ICT) in recent decades. The home service offered by the smart devices helps the users in utilize maximized level of comfort for the objective of improving life quality. As the user and smart devices communicate through an insecure channel, the smart home environment is prone to security and privacy problems. A secure authentication protocol needs to be established between the smart devices and the user, such that a situation for device authentication can be made feasible in smart home environments. Most of the existing smart home authentication protocols were identified to fail in facilitating a secure mutual authentication and increases the possibility of lunching the attacks of session key disclosure, impersonation and stolen smart device. In this paper, Secure Supervised Learning-based Smart Home Authentication Framework (SSL-SHAF) is proposed as are liable mutual authentication that can be contextually imposed for better security. The formal analysis of the proposed SSL-SHAF confirmed better resistance against session key disclosure, impersonation and stolen smart device attacks. The results of SSL-SHAF confirmed minimized computational costs and security compared to the baseline protocols considered for investigation.*

## KEYWORDS

*Supervised Learning, Smart Home, Authentication Framework, Contextual Information, and Mutual Authentication.*

## 1. INTRODUCTION

In the recent past, the information shared among IoT devices has exponentially increased with an indispensable increase in the essentiality of security [1]. In specific, security is an extensive domain [2]. This security domain is categorized into four classes that include service availability protection, availability, alteration prevention against exchanged information, integrity, provisioning security over communication channels established between communicating parties, confidentiality, claimed identity validation and authenticity. In specific, authenticity is the first important class of security over which the other classes of security are constructed [3]. Moreover, the approaches of authentication are classified into physiological, object-based, and knowledge-based depending on the dimension used for provisioning security [4]. However the authentication approaches possess limitations, since the knowledge-based approaches necessitate the remembrance of credentials associated with 1authentication that includes username and password which are highly susceptible to attacks. In particular, the key challenge in the SH environment wholly emphasizes on enabling secure and effective user authentication that confirms that the command is certainly dispensed by the genuine user or device [5]. A safe distant authentication mechanism using Supervised Learning (SL) that allows only genuine users to gain admittance to smart devices is indispensable. The attacker has the likelihood of achieving access to the home so as to perform criminal activities like theft or attack, when he efficaciously negotiates the smart door for accepting the false command [6].

A varied amount of authentication protocols is available for focusing on the production of safe session keys that can be used for the production of authentic channels [7]. IoT in an SH does not own tolerable storage and computation capacity for assisting in the application of the prevailing protocols focusing on exhaustive cryptographic algorithms. Furthermore, the existing authentication schemes [8] demand an increased amount of user interference based on provisioning and configuration. Many devices have limited access, thus requiring primary configuration to protect from stealing, meddling and possible forms of complete negotiation throughout their lifetime [9]. Further, password-based results are not the only probable solutions as the passwords are likely to be easily broken. Additionally, many IoT devices do not act as interface for password endorsement. Despite varied user authentication methods propounded for achieving remote access, they do not focus on the past transactions and awareness of physical context [10]. These features are necessary in an authentication scheme for overcoming Mirai kind of attacks. Furthermore, the prevailing protocols are found to be complex and not preferably secure for precluding attacks on the resource-controlled devices in a smart home environment.

## 1.1. Motivation

The algorithms that use supervised learning generally operates using labelled datasets. It is mainly used for detecting the locations, adaptive filtration, channel estimation, and spectrum detection during the access of smart home devices. These supervised learning approaches can be categorized into classification and regression types. Some of the most significant classification algorithms that can be used in smart home environments are decision trees, random forest, naïve Bayes, and Support Vector Machine (SVM, etc. On the other hand, logistic regression and polynomial methods are considered as the regression methods which can be ideally utilized in the smart home environments. In specific, these supervised algorithms pertain to the category of instance-oriented algorithms for predicting the output depending on the model learned from each new observation identified from the smart home scenario. For IoT-based smart home environments, supervised learning algorithms such as naïve Bayes, DT and SVM can be utilized. For example, the constraints with non-linear properties can be used for a model of solution that could be obtained from SVMs. However SVMs are highly insufficient for massive datasets. When massive datasets are used, random forest algorithms can be applied to facilitate maximized accuracy degree. It also requires minimized amount of prediction. But the training process of random forest algorithm incurs minimized time compared to NB and SVM algorithms. Moreover, supervised learning models are used for thwarting DDoS attacks and detect intrusion in the IoT network in the cloud and the layer of communication.

In this paper, a secure Supervised Learning-based Smart Home Authentication Framework (SSL-SHAF) is proposed for achieving potential mutual authentication to attain better security. This SSL-SHAF framework is implemented based on three different supervised learning strategies that incorporate and strategies together in a more contextual way. The formal investigation of the proposed SSL-SHAF scheme aided in offering better resistance against session key disclosure, impersonation and stolen smart device attacks. The experimental validation of the proposed SSL-SHAF is also conducted using computational costs and computational cost to compare it with the competitive baseline protocols considered for investigation.

## 2. RELATED WORK

A secure smart home mutual authentication framework using multi-factor was proposed for utilizing transitory identities towards better security provision in smart home environments [11]. This mutual authentication framework was proposed for confirming the authentication of users with the controllers in the smart building environment in a more untraceable, unlinkable and anonymous manner. It was proposed with the capability of preventing the issue of clock synchronization. It was also proposed for resisting quantum computing attacks to prevent the credentials of the users in the smart home environment. The model checks and informal analysis of this smart home mutual authentication framework confirmed minimized computations cost and communication overhead with maximized capability in thwarting security attacks in the smart home scenario. Another, smart home authentication framework was proposed for guaranteeing high-level security using the single low-entropy human memorable password for securing smart home scenarios [12]. It was proposed as a key confirmation protocol that utilized the merits of mutual authenticated key agreement, password, and threshold for securing smart home environment. It was proposed for mitigating l−1 smart home devices for handling the impacts of the adversary in a more scalable and robust scheme. It was proposed for designing password-intensive settings for ensuing end-to-end security depending on the selected IoT devices during the process of user authentication. It was proposed for facilitating session key secrecy and mutual authentication with respect to the user and the device manager. It confirmed minimized computation and communications costs independent the number of users authenticated in the smart home environment.

An authentication Framework using a cancellable Biometric System (CBS) was proposed for securing smart home user credentials from exploitation [13]. This CBSAF was proposed as a biometric protection approach for performing the operation of authentication provided at the transformation or distortions degree over the level of the features or signals. It was proposed for provisioning potential environment that makes it more suitable and deployable in real time settings that improves the maximized accuracy, minimized overhead without influencing the security of the sensitive biometric templates. The experimental and theoretical investigations of CBS confirmed better minimized equal error rate on par with the existing works of the literature. It was identified to incurs less time in order to make it more ideal for IoT environments. In addition, a Context-aware smart home authentication framework (CASHAF) was proposed for ensuring security based on the contextual information derived with respect to the patterns of users' access behaviour, request time, calendar, profile and location of the user [14]. This authentication approach used the derived confidential information for making concluding decisions for validating the access requests for accepting or rejecting the request access of the users in the smart home scenario. It was proposed for facilitating security in a more adaptive manner without the intervention of the user. The core objective of this CASHAF completely concentrates on the process of determining the continuous authentication model used by mobile users for the purpose of accessing smart home devices [15]. It was proposed as an application for IoT depending on the use of classical credentials that include potential contextual information. It was also proposed a multifactor authentication attribute that integrates the merits of context information and static credentials.

## 3. PROPOSED SECURE SUPERVISED LEARNING-BASED SMART HOME AUTHENTICATION FRAMEWORK (SSL-SHAF)

In this section, the detailed view of the proposed SSL-SHAF with the steps involved during the smart authentication process.

The primary goal of the proposed SSL-SHAF framework concentrates on the process of attaining a continuous authentication mechanism for mobile user clients essential for accessing smart home services (devices). This SSL-SHAF framework utilizes contextual information using supervised learning depending on the utilized classical credentials. This SSL-SHAF framework is implemented as a use-case scenario for facilitating services towards user authentication in the smart home environment. It derived contextual information from the smart home resources deployed in the environment for examining it in the real time. It utilized multifactor attributes that includes the integration of context information and static credentials for strengthening the process of authentication. It facilitated better decision-making process through the inclusion of different authentication attributes that attributes to assigned level of confidence associated with each smart device interacting in a specific implemented environment. It adopted different levels of confidence to each of its registered users in the smart home scenario based on the estimated threshold level of access. This confidence level proves wide options for the smart home owner in feasibly deriving available factors depending on the views that could be possibly determined in the smart home scenario. In specific, contextual information utilized in the framework includes historical information such as logs and access patterns, calendar of the users, Bluetooth and IP information of the location, and profile of the users including identifier, age, and name.

The proposed SSL-SHAF framework comprises of the following components explained as follows.

a) **Clients (Users):** The user represents any individual entity or a person who attempts to access of the protected devices or services that could be possibly facilitated through that device.
b) **Home devices:** These devices represent any equipment that interacts in the IoT environment through the smart lock, thermostat and IP camera which can be potentially accessed wirelessly by the clients through the aid of the gateway.
c) **Database of the users:** It comprises of different tables that need to be updated in the smart home environment in the form of encryption with a calendar, user profiles, usage patterns and access control policies.
d) **Core gateway:** It is the intermediator entity that exists between the connected home device and the clients. It plays an anchor role in protecting the user access to the smart home devices and achieving the process of authentication. It is responsible for gathering the necessitated contextual information. It aids in verifying whether the access request satisfies the requirement in a more pre-defined manner. It represents an application that gets executed over a small, single-board computer named Raspberry Pi. Moreover, the access policies need to be controlled through the gateway.
e) **Bluetooth Sensor:** It is the real-time sensor embedded in the Raspberry Pi for facilitating information collection which pertains to the information associated with the proximity of the users.

The use case scenario considered during the implementation of the proposed SSL-SHAF framework comprises the following steps, i) stage of registration, ii) stage of verification, iii) stage of login, iv) stage of utilization and v) stage of authentication. In the initial stage of registration, the users need to provide specific details that include the calendar schedule and preferences of the users. In the stage of verification, the user termed as the home owner is responsible for activating the user accounts and reviewing the user registration. In the stage of login, the users initially provide their classical credentials for achieving the maximized authentication level that is related to the smart devices based on the user profiles. However, when the user external to the home environment wants to access the services of the framework, proper access with restriction is provided based on the roles and predefined policies provided by the home owner. In the stage of smart home service utilization, the user performs access to the smart equipment's with the help of web GUI. Moreover, the gateway continuously confirms the access of the requested users depending on the information derived in different contexts includes history of access, location and calendar. Finally, the access level of users in the authentication stage is achieved based on the computation of contextual information integrated with estimated different levels of confidence.

The proposed SSL-SHAF framework inherits the merits of three different supervised learning-based smart home authentication mechanism as detailed as follows.

### 3.1. Merkle Hash Tree based Enhanced Strong Discrete Hashing Function Authentication Scheme (MHT-ESDHF-AS)

MHT-ESDHF-AS approach is proposed based on transaction history and physical context for guaranteeing security. It alleviates the challenges of clock synchronization and does not involve a verification table during authentication [14]. It is proposed as a supervised hashing method that prevents the issues of huge quantization errors and sub-optimal Binary Codes (BCs). It is proficient in learning BCs and fighting against transmission loss as the information used for verification includes the authentication data itself. The computation and communication costs of the proposed scheme is found to be lesser than the standard authentication mechanisms taken for investigation. The formal, informal and model checking-based security examination of the propounded MHT-ESDHF-AS is also found to be better than the standard authentication mechanisms designed for SH environments. The proposed MHT-ESDHF-AS is a secure and lightweight scheme designed for attaining mutual authentication and key agreement using the MHT-ESDHF supporting trusted communication for SH applications. It incorporates the transaction history and information about the context for improving the security level for remote access of IoT-based SH environments. It overcomes the issue related to clock synchronization that is common in timestamp-based authentication and also avoids the use of averification table.

### 3.2. Supervised Hash Signature using Dynamic Forest of Random Subsets (SHS-DFRS) Mechanism

This SHS-DFRS is propounded as an anonymous authentication approach based on past transactions and context awareness during user authentication in SHs. This propounded SHS-DFRS mechanism simultaneously performs the signed production of arbitrary subsets as every element of a signature is involved in the production of consecutive arbitrary elements in authentication. It uses the power of exposed secret keys by binding the packages to arbitrary subset production and signature. The SHS-DFRS is designed as a lightweight and trusted user authentication approach that is extremely good for a SH[17]. It combines the advantages of exposed secret keys depending on chaining that incorporates the signature and subset production. It includes a principal SL scheme that permits only the owner of the message to possibly produce a arbitrary subset so as to avoid malicious invaders from accessing smart devices. The proposed SHS-DFRS mechanism is a secure and lightweight key management and authentication protocol for distant access of IoT supported SHs. It includes key generation (Secret and public keys, and quantum SHT) based on past transactions attained using SL for applying exploited security. It is based on the advantages of Dynamically Obtained Random Subset (DORS)-based signature production and confirmation of the produced signature to support easiness in an application.

### 3.3. Supervised Learning-based Discrete Hash Signature Authentication (SLDHSA)

This supervised authentication scheme facilitates the benefits of similarity learning using the characteristics of symmetric cryptosystems and discrete hash signatures. It is proposed for addressing security and authentication through the assignment of unique addressing to the devices that aids in constructing the smart home scenario, It includes distinct identification based on the updated unique 64-bit interface identifier generated during the process of authentication. It generated discrete hash signature based on the 64-bit interface identifier exchanged between the user smart card, edge server and home server. It always stores the user request packet in the encapsulated format such that the least significant 64 bits are available in the IPv6 packet format. In specific, the edge server is responsible for the reception of the user request for decapsulating the unique identifier. It is proposed for verifying whether the identifier stored in the database of the edge server and user smart card are similar during the process of authentication. It also utilizes the merits of home server using the registration process for storing the distinct identifier in the database to forward it and update the information of the smart card. It adopts six important phases that include initialization, addressing, registration, login authentication, session agreement and password update. In addition, the overall framework diagram of the proposed SSL-SHAF framework is depicted in Figure 1.

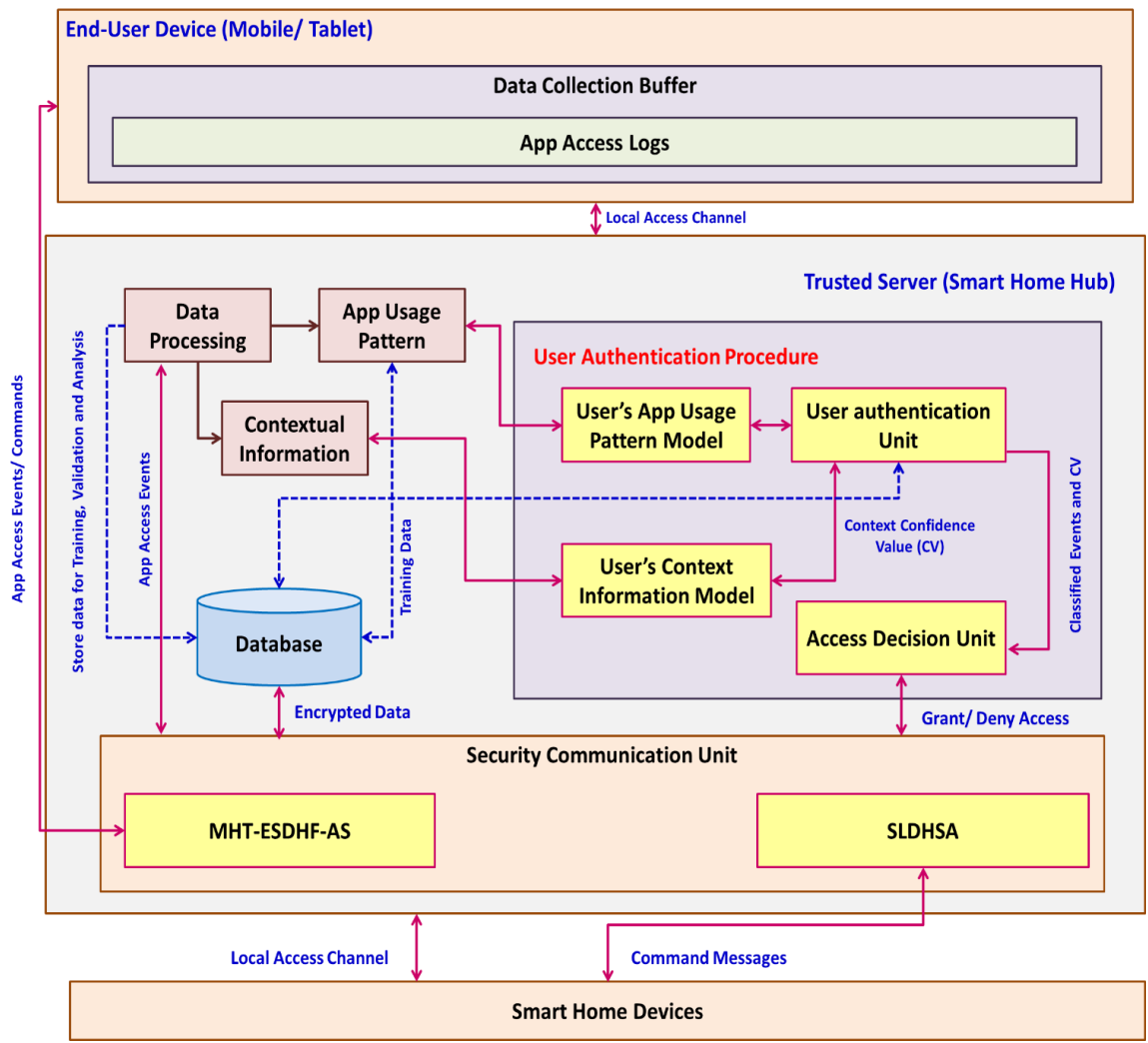

Figure 1: Framework diagram of the proposed SSL-SHAF

## 4. RESULTS AND DISCUSSION

The performance of the proposed SSL-SHAF framework and the competitive CASHAF [15] and CBSAF [16] methods framework is validated based on Internet Access Time and Local Access Time measured in milliseconds.

Initially, Table 1 and 2 demonstrates the Internet Access Time and Local Access Time incurred by the proposed SSL-SHAF framework and the competitive framework used for investigation.

Table 1: SSL-SHAF Framework-Internet Access Time and Local Access Time under individual algorithm enforcement

| Utilized parameter | Internet access time (milliseconds) | | | Local Access Time (milliseconds) | | |
|---|---|---|---|---|---|---|
| | SSL-SHAF | CASHAF[14] | CBSAF[13] | SSL-SHAF | CASHAF[14] | CBSAF[13] |
| Proximity (Bluetooth-based location | 86 | 94 | 99 | 7 | 11 | 15 |
| Access of Calendar | 96 | 112 | 134 | 13 | 19 | 26 |
| Network (IP address-based location) | 92 | 96 | 103 | 13 | 17 | 23 |
| Username and password (knowledge-based credentials) | 93 | 111 | 123 | 14 | 19 | 24 |
| No authentication | 83 | 91 | 98 | 6 | 10 | 15 |

Table 2: SSL-SHAF Framework-Internet Access Time and Local Access Time under integration of algorithms

| Utilized parameter | Internet access time (milliseconds) | | | Local Access Time (milliseconds) | | |
|---|---|---|---|---|---|---|
| | SSL-SHAF | CASHAF[14] | CBSAF[13] | SSL-SHAF | CASHAF[14] | CBSAF[13] |
| Bluetooth and IP Address-based location | 83 | 90 | 97 | 11 | 14 | 17 |
| Bluetooth and IP Address-based location with access of Calendar | 92 | 06 | 103 | 13 | 17 | 23 |
| Bluetooth and IP Address-based location with access of Calendar and knowledge-based credentials | 93 | 98 | 111 | 13 | 16 | 19 |
| No authentication | 82 | 88 | 94 | 5 | 8 | 11 |

The above demonstrated results from Table 1, clearly confirmed that the proposed SSL-SHAF is performing better than the competitive CASHAF [15] and CBSAF [16] methods independent of the process of integrating attributes or individually selecting contextual information. As the expected, the integration of all the methods is considered to be comparatively higher than the no authentication condition. But a minimum degree of overhead is realized with the request level related to the methods used for authentication. Moreover, the response time is identified to be influenced more due to the proximity, calendar access and knowledge-based credentials.

In the second level of investigation, the proposed SSL-SHAF and the competitive CASHAF [15] and CBSAF [13] methods are compared based on the security factors of integrity, availability, and authentication and presented in Table 3.

Table 3: Performance comparison of the proposed SSL-SHAF with respect to security factors

| Security factors considered for evaluation | Schemes under comparison | | |
|---|---|---|---|
| | SSL-SHAF | CASHAF[14] | CBSAF[13] |
| Authentication | Strong | Strong | Strong |
| Integrity | Strong | Moderate | Moderate |
| Availability | Strong | Strong | Weak |

The above-mentioned results depicted in Table 3 demonstrated that the proposed SSL-SHAF framework is competent enough in guaranteeing the factors of security such as authentication, Integrity and Availability on par with the CASHAF [15] and CBSAF [16] methods used for comparison. Further, the performance of the proposed SSL-SHAF and the competitive CASHAF [15] and CBSAF [16] authentication frameworks are compared based on response time, computational overhead, and communicational overhead. Figures 2, 3 and 4 depict the response time, computational overhead and communicational overhead incurred by the proposed SSL-SHAF and the competitive CASHAF [15] and CBSAF [16] authentication frameworks with respect to number of users. The results confirmed that the proposed SSL-SHAF has potential enough in minimizing the response time as contextual information based on supervised learning is derived during the authentication process. In particular, the computational cost of the proposed SSL-SHAF framework is minimized, since the number of messages used for authentication is minimal as they integrate the factors of Bluetooth and IP Address-based location with access of Calendar and knowledge-based credentials together in a more contextual manner. Moreover, the proposed SSL-SHAF framework also reduced the storage overhead on par with the baseline frameworks of comparison as they contextually adopt only one supervised learning-based user smart home authentication at a single instant of time.

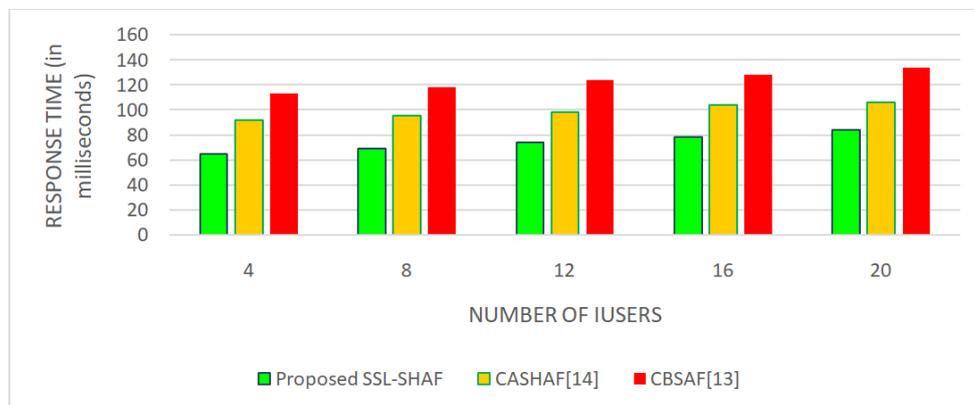

Figure 2: Proposed SSL-SHAF-Response timeunder different users

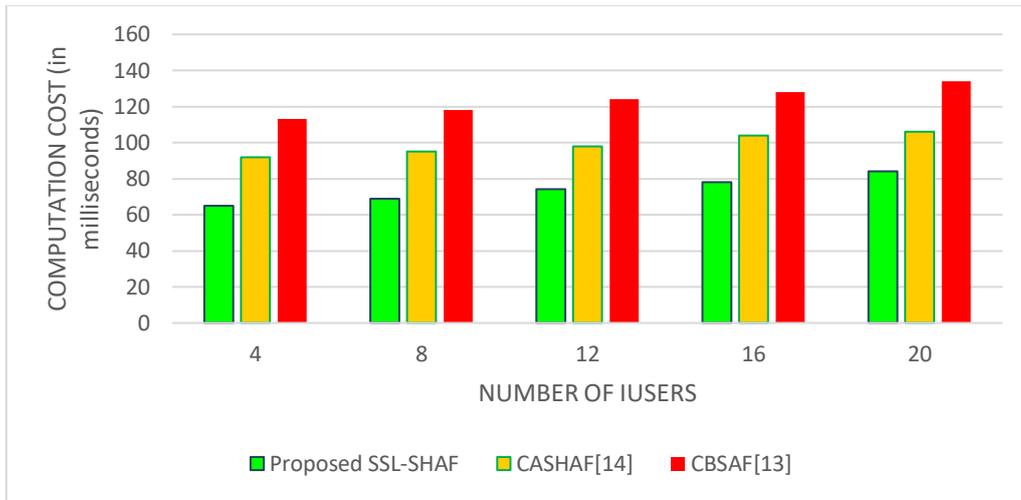

Figure 3: Proposed SSL-SHAF-communicational cost under different users

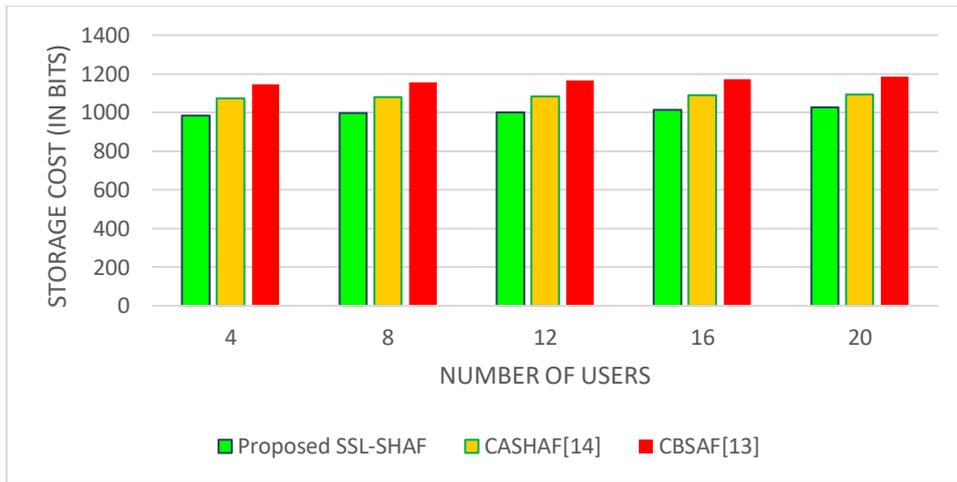

Figure 4: Proposed SSL-SHAF-Storage Costs (in bits) incurred under different users

Thus, the SSL-SHAF framework, on average minimized the response time by 19.21%, computation cost by 21.84% and storage costs by 18.72%, independent to the users requisitioning for anauthentication process.

In addition, the proposed SSL-SHAF framework is compared with the baseline MHT-ESDHF-AS, SHS-DFRS and SLDHSA with respect to privacy preservation degree under different users considered for evaluation. Figure 5 demonstrates the privacy preservation degree achieved by the proposed SSL-SHAF framework on par with the compared MHT-ESDHF-AS, SHS-DFRS and SLDHSA under different number of users.

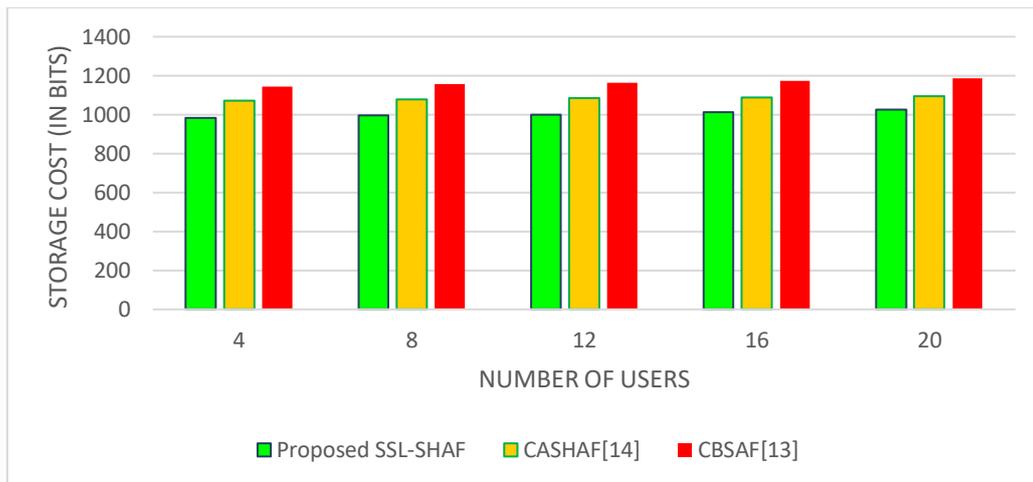

Figure 5: Proposed SSL-SHAF- Privacy preservation degree achieved under different users

The results from Figure 5 confirmed that the proposed SSL-SHAF framework, on average improved the privacy preservation degree by 18.21% and 20.18%, better than the benchmarked authentication process.

## 5. CONCLUSION

The proposed SSL-SHAF achieved robust and secure mutual authentication through contextual derivation of parameters associated with access control. It facilitated better resistance against session key disclosure, impersonation and stolen smart device attacks. It integrated the benefits of MHT-ESDHF-AS, SHS-DFRS and SLDHSA in a more contextual manner for guaranteeing security against attacks that could be launched in the smart home scenario. The results of SSL-SHAF confirmed minimized computational costs and security compared to the baseline protocols considered for investigation. The investigational endorsement of the propounded SSL-SHAF framework confirmed the decrease in the computational overhead by 6.74%, 7.92% and 9.568% in contrast to the standard schemes. The storage and communication costs of the propounded SSL-SHAF also found to be reasonably reduced when compared to the standard authentication schemes in a SH. As the part of future scope, it is decided to formulate a semi-supervised learning-based mutual authentication framework and compare it with the proposed supervised learning framework.

## CONFLICTS OF INTEREST

The authors have no conflicts of interest to declare. All co-authors have seen and agree with the contents of the manuscript and there is no financial interest to report.

## AUTHORS

**K. Swapna Sudha** received M.Tech degree from Jawaharlal Nehru Technological University, Anantapur, Andhra Pradesh, India. Currently, she is a research scholar in the School of Computer Science Engineering and Information Systems, Vellore Institute of Technology, Vellore, Tamilnadu, India and pursing her Ph.D. degree in the field of Cyber Security. Her main area of research includes Network Security, Cyber Security and Internet of Things. 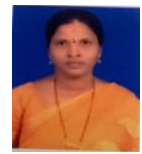

**Dr. N. Jeyanthi** received her Ph.D. degree in Cloud Security from VIT University, Vellore, Tamilnadu, India. She is an Associate Professor in VIT, Vellore for School of Computer Science Engineering and Information Systems. Her research work was funded by Department of Science and Technology, Govt. of India. She has authored and co-authored over 62 research publications in peer-reviewed reputed journals and 30 conference proceedings. Her entire publications have been cited over 367 times (Google Scholar). The latest Google h-index of his publications is 12 and i10 index is 13. Books 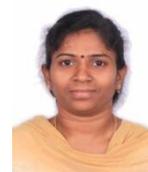 and book chapters were also added to her research contribution. She has served as the program committee member of various international conferences and reviewer for various international journals. She has been honoured by VIT as an active researcher for four consecutive years. Her current areas of interest include IoT, Cloud, and Big Data security.

**Celestine Iwendi** (Senior Member, IEEE) received the master's degree in communication hardware and microsystem engineering from Uppsala University, Uppsala, Sweden, in 2008, and the Ph.D. degree in electronics from the University of Aberdeen, Aberdeen, U.K., in 2013. He is currently a Senior Lecturer with the School of creative Technologies, University of Bolton, Bolton, U.K. He is also a Visiting Professor with Coal City University Enugu, Enugu, Nigeria. He is a highly motivated and hardworking researcher. He has authored or co-authored the book Wireless Sensor Network Security and more than 100 publications. His research interests include the Internet of Things, machine learning, artificial intelligence, and wireless sensor networks. He is a fellow of the Higher Education Academy, U.K. He is also a Board Member of IEEE Sweden Section. 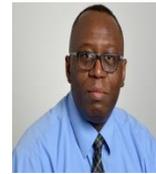